\documentclass[]{iopart}
\usepackage{graphicx}
\usepackage{iopams}

\expandafter\let\csname equation*\endcsname\relax

\expandafter\let\csname endequation*\endcsname\relax

\usepackage{amsmath}

\usepackage{color}
\usepackage{bm} 
\usepackage{bbm, dsfont} 
\usepackage{subfigure}
\usepackage{mathrsfs}
\usepackage{verbatim}
\usepackage{cite}
\usepackage{amsmath} 

\begin{document}

\title{Crossover from a delocalized to localized atomic excitation in an atom-waveguide interface}

\author{H. H. Jen}
\address{Institute of Atomic and Molecular Sciences, Academia Sinica, Taipei 10617, Taiwan}
\ead{sappyjen@gmail.com}
\author{J.-S. You}
\address{Department of Physics, National Taiwan Normal University, Taipei 116, Taiwan}
\def\q{\mathbf{q}}
\renewcommand{\k}{\mathbf{k}}
\renewcommand{\r}{\mathbf{r}}
\newcommand{\parallelsum}{\mathbin{\!/\mkern-5mu/\!}}
\def\p{\mathbf{p}} 
\def\R{\mathbf{R}}
\def\bea{\begin{eqnarray}}
\def\eea{\end{eqnarray}}
\begin{abstract}
An atom-waveguide system, which presents one of the quantum interfaces that enable strong couplings between light and atoms, can support tightly-confined guided modes of light. In this distinctive quantum interface, we theoretically investigate the crossover from a delocalized to localized atomic excitation under long-range dipole-dipole interactions and lattice disorders. Both localization lengths of the excitation distributions and power-law scalings of dissipative von Neumann entanglement entropy show signatures of this crossover. We further calculate numerically the level statistics of the underlying non-Hermitian Hamiltonian, from which as the disorder strength increases, the gap ratio decreases and the intrasample variance increases before reaching respective saturated values. The mean gap ratio in the deeply localized regime is close to the one from Poisson statistics along with a relatively large intrasample variance, whereas in the nondisordered regime, a significant level repulsion emerges. Our results provide insights to study the non-ergodic phenomenon in an atom-waveguide interface, which can be potentially applied to photon storage in this interface under dissipations.         
\end{abstract}
\maketitle
\section{Introduction}

An atom-waveguide system \cite{Balykin2004, Sague2007, Morrissey2009, Vetsch2010, Goban2012, Luxmoore2013, Yala2014, Sollner2015, Solano2017_2}, one of the nanophotonics platforms \cite{Kien2005, Kien2008, Tudela2013, Lodahl2015, Kumlin2018, Chang2018}, enables strong and confined light-matter interactions, which is essential for an efficient and controllable quantum interface \cite{Hammerer2010}. This strong coupling can be achieved from the guided modes of evanescent waves \cite{Bliokh2015} carried by the one-dimensional (1D) optical nanofiber. Recently, superradiance \cite{Dicke1954, Lehmberg1970, Gross1982} has been experimentally demonstrated in two atom clouds \cite{Solano2017} separated far beyond the transition wavelength, and single collective excitation can be stored and emitted on demand via coupling to a nanoscale waveguide \cite{Corzo2019}. The former presents the feature of long-range couplings in the 1D dipole-dipole interactions mediated by a nanofiber, while the latter shows the capability to process nonclassical quantum states of single photons. These achievements can lead to scalable interconnection nodes between light and atoms, which are essential to large-scale quantum network \cite{Kimble2008}. 

In the atom-waveguide system under strong coupling regime, spin dimers \cite{Stannigel2012, Ramos2014, Pichler2015} can be formed from chiral dissipations \cite{Lodahl2017}, photon transport \cite{Jen_transport_2019, Needham2019} can be strongly correlated \cite{Mahmoodian2018}, and subradiant states \cite{Albrecht2019} and their dynamics \cite{Jen2020_subradiance, Jen2020_subradiance2} emerge owing to the collective light-atom couplings. This collective interaction also promises an enhanced performance of photon storage \cite{Garcia2017}, which marks the advantage of quantum resource that can be harnessed from the 1D dipole-dipole interactions in the atom-waveguide interface. Under a weakly driven-dissipative setting \cite{Jen2020_PRR}, the dynamics of atomic excitations is predicted to sustain for a long time, resembling the persistent oscillations of the dipolar spin impurities in diamond \cite{Choi2017}, and shows clumps of atomic excitations as time evolves. This can be related to the constrained many-body scars \cite{Turner2018, Voorden2020} that break the ergodicity of the system, where slow thermalization further leads to strong localization of an atomic excitation \cite{Jen2020_disorder} when the atoms' position disorders are introduced as in Anderson model \cite{Anderson1958}.  

Anderson-like localization under dissipations and 1D dipole-dipole interactions is less explored owing to the complexity arising from the strong atom-atom correlations in the atom-waveguide system. Clear quantitative measures to determine the localization transition under dissipations are also lacking. Here we use the level statistics to present the crossover from a delocalized to localized atomic excitation in an atom-waveguide interface, along with other insightful but nondeterministic measures of localization length and von Neumann entanglement entropy. First we formulate the effective non-Hermitian Hamiltonian under single excitation space. Then we show that the localization lengths determined from the excitation distributions change in time, and the power-law fitting of von Neumann entanglement entropy indicate the onset of localization only qualitatively owing to the system's dissipative nature. Next we introduce the level statistics, including the gap ratios and intrasample variance, to analyze the crossover to the excitation localization, which converges as we increase the number of atoms. We also present an interaction-driven delocalization as we tune the interparticle distance of a periodic atomic array. Our results offer opportunities to study the nonergodic phenomena in an atom-waveguide interface under dissipations by tailoring competing atom-atom interactions and disorder strengths.    

\section{Effective theoretical model}

\begin{figure}[t]
\centering
\includegraphics[width=10.0cm,height=7.5cm]{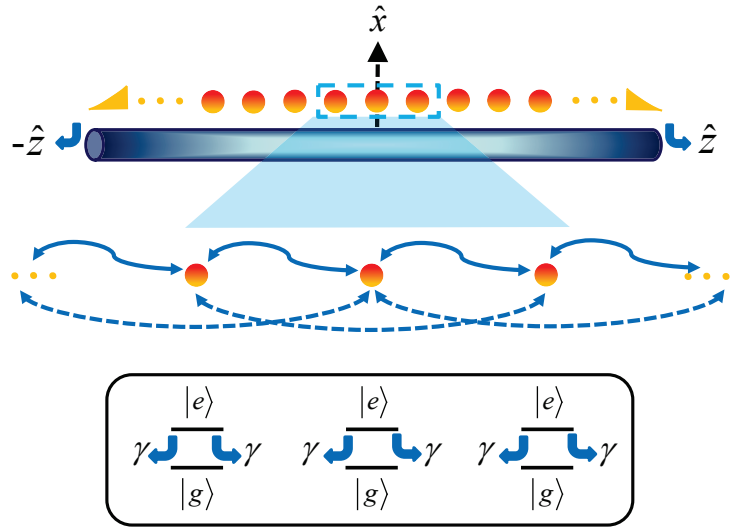}
\caption{Schematic plot of an atom-waveguide interface. A periodic array of atoms aligning close to the nanofiber couple with each other with one-dimensional dipole-dipole interactions. This long-range interaction emerges via the guided modes supported by the waveguide along $\hat z$. A zoom-in plot of a dashed box of three atoms presents the effect of dipole-dipole interactions which couple the nearest-neighbor atoms (thick double arrows) and beyond (dashed double arrows). Each atom is a two-level quantum emitter ($|g\rangle$ and $|e\rangle$ as the ground and the excited states, respectively) which decays with an intrinsic rate of $\gamma$ to either sides of the waveguide.}\label{fig1}
\end{figure}

We consider an atomic array of $N$ two-level quantum emitters with $|g\rangle$ and $|e\rangle$ for the ground and excited states respectively. This periodic atomic array positioned close to a waveguide or nanofiber, as schematically shown in figure \ref{fig1}, presents one of the strongly interacting quantum interfaces that allow exchanges of light quanta and atomic excitations \cite{Corzo2019}. Owing to the the guided modes supported in the waveguide, the atoms couple to the evanescent fields near the waveguide, which results in the light-induced dipole-dipole interactions effectively \cite{Solano2017}. This interaction builds up strong atom-atom correlations at long distances where higher dimensional quantum optical interfaces are unable to match. 

For an atom-waveguide system in an interaction picture, the dynamics of the density matrix $\rho$ can be described by the effective theoretical model \cite{Tudela2013, Pichler2015} in a Lindblad form ($\hbar$ $=$ $1$), 
\bea
\frac{d \rho}{dt}=-i[H,\rho]+\mathcal{L}[ \rho],\label{rho}
\eea 
where the respective coherent and dissipative parts are  
\bea
H =&& \gamma\sum_{\mu\neq\nu}^N\sum_{\nu=1}^N \sin\left(k_s|r_\mu-r_\nu|\right)\sigma_\mu^\dag\sigma_\nu, \label{H}
\eea
and
\bea
\mathcal{L}[\rho]=&&-\gamma\sum_{\mu,\nu=1}^N \cos\left[k_s(r_\mu-r_\nu)\right]\left(\sigma_\mu^\dag \sigma_\nu \rho + \rho \sigma_\mu^\dag\sigma_\nu -2\sigma_\nu \rho\sigma_\mu^\dag\right).
\eea
The decay rate $\gamma$ is defined as $2|dq(\omega)/d\omega|_{\omega=\omega_{eg}}g_{k_s}^2L$ \cite{Tudela2013} with the group velocity $|d\omega/dq(\omega)|$ for the wave vector $q(\omega)$ at the resonant frequency of two-level transition $\omega_{eg}$, the coupling strength $g_{k_s}$ for the wave vector $k_s$ in the guided mode, and the quantization length $L$. The raising and lowering dipole operators are respectively $\sigma_\mu^\dag$ $\equiv$ $|e\rangle_\mu\langle g|$ and $\sigma_\mu$ $=$ $(\sigma_\mu^\dag)^\dag$, and the atoms can be ordered as $r_\mu$ $<$ $r_{\mu+1}$ for $\mu$ $=$ $[1,N-1]$.   

Equation (\ref{rho}) in general describes the spin-exchange process \cite{Dicke1954} (two-level quantum emitters equivalent to a spin-1/2 system) with long-range complex hopping terms under dissipations. To investigate the excitation localization as in Anderson's model, we assume the central atom is flipped to be spin up initially. This spin would then relax and experience multiple scatterings through the atoms under long-range dipole-dipole interactions before exiting the whole array, in contrast to the conventional Anderson's absence of spin diffusion \cite{Anderson1958} which assumes short-range interactions and lossless environment. Since single spin is flipped in the first place, it is legitimate to focus on single excitation space $|\phi_\mu\rangle$ $=$ $\sigma_\mu^\dag |g\rangle^{\otimes N}$ along with the probability amplitudes $a_\mu(t)$. Within this subspace with the conserved number of one spin up in total and under the added phase disorders $W_\mu$, we obtain the system's time evolution, $|\Phi(t)\rangle$ $=$ $\sum_{\mu=1}^N a_\mu(t)|\phi_\mu\rangle$, governed by the Schr\"{o}dinger's equation, 
\bea
i\frac{\partial}{\partial t}|\Phi(t)\rangle=H_{\rm n}|\Phi(t)\rangle,\label{nonH}
\eea  
where a non-Hermitian Hamiltonian $H_{\rm n}$ under disorders reads 
\bea
H_{\rm n}=&&-\gamma\sum_{\mu\neq\nu}^N\sum_{\nu=1}^N e^{-i(\mu-\nu)\xi-i(W_\mu-W_\nu)}\sigma_\mu^\dag\sigma_\nu -i\gamma\sum_{\mu=1}^N\sigma_\mu^\dag\sigma_\mu.
\eea
The dimensionless interparticle distance $\xi$ $\equiv$ $k_s |r_{\mu+1}-r_{\mu}|$ characterizes the effect from dipole-dipole interactions, and the onsite phase disorders $W_\mu$ are uniformly distributed within $[-w,w]$ where $w/\pi$ $=$ $[0,1]$. $W_\mu$ can arise and can be controlled from the position fluctuations around $\xi$ when the atoms are loaded to shallow optical lattices. 

We note that equation (\ref{nonH}) without disorder has been used to investigate the subradiant dynamics \cite{Jen2020_subradiance, Jen2020_subradiance2}, the steady-state phase diagram under a weakly-driven condition \cite{Jen2020_PRR}, and strong localization of atomic excitation in a chirally-coupled atomic chain \cite{Jen2020_disorder}. Here we focus on the eigenspectrum of an atom-waveguide system with reciprocal couplings, from which its level statistics can be used to uncover the distinctive delocalization to localization crossover in an atom-waveguide interface as the disorder strength increases. 

\section{Localization lengths and von Neumann entanglement entropy}

The localization length determined from the spread of the atomic excitation can be an indicator of excitation localization. When this spread touches the system boundary, we can qualitatively claim the onset of delocalization from the localization side \cite{Choi2016}. However, it is not straightforward to apply this measure to the atom-waveguide system under dissipations, since the localization length is modified as time evolves. In addition to the localization length characterization, the entanglement entropy, which quantifies the global configuration of how spin diffuses through the atomic chain, provides an alternative way to identify the localization phenomena. It has been postulated that a logarithmic growth of entanglement entropy in time \cite{Znidaric2008, Bardarson2012, Schreiber2015, Lukin2019, Kiefer2020} is characteristic of the many-body localized phase \cite{Abanin2019}. In contrast to the growth of entanglement, we have a decaying one owing to the relaxation of the atomic excitation, which shows a small power-law scaling in time when the system is evolving through a strong disorder.   
 
\begin{figure}[t]
\centering
\includegraphics[width=12.0cm,height=6cm]{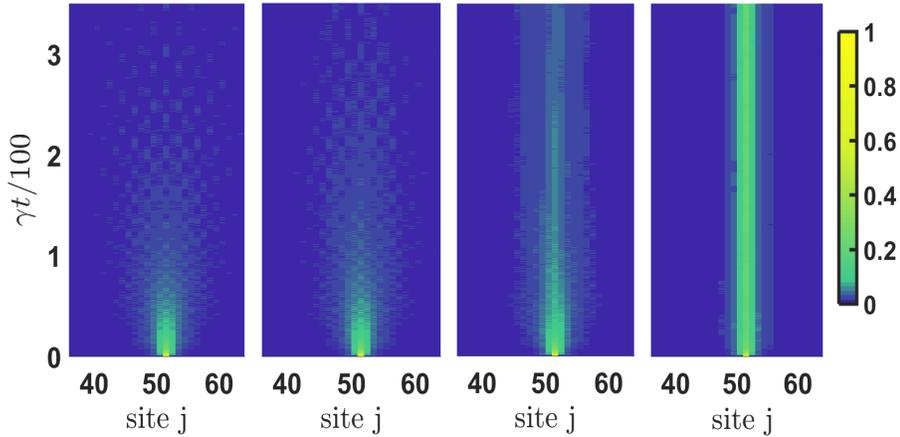}
\caption{Atomic excitation ($\langle P_j(t)\rangle$) localization. The central atom is initially excited at the $51$th site of the periodic atomic array with $N$ $=$ $101$ and $\xi$ $=$ $\pi/8$. From the left to the right panels, a crossover from the excitation delocalization to localization is presented as the disorder strengths increase from $w/\pi$ $=$ $0$, $0.01$, $0.025$ to $0.1$. The presented results here and throughout the paper are converged after averaging over $200$ realizations of disorders.}\label{fig2}
\end{figure}

To see how the atomic excitation diffuses as time evolves, we numerically solve equation (\ref{nonH}) with the initially flipped spin that $a_{(N+1)/2}(0)$ $=$ $1$ for an odd $N$. Under single excitation space, the system dynamics can be solved directly from   
\bea
\dot{a}_\mu(t)= &&-\gamma\sum_{\nu\neq\mu}^N e^{-i(\mu-\nu)\xi-i(W_\mu-W_\nu)}a_\nu(t)-\gamma a_\mu(t). \label{coupling}
\eea
In figure \ref{fig2}, we show the emergence of atomic excitation localization as the disorder is tuned from a weak to a moderate strength, where the spatial and time evolutions of $\langle P_j\rangle$ $\equiv$ $\langle|a_j|^2\rangle$ with an ensemble average $\langle\cdot\rangle$ are plotted. For the case without or with a weak disorder, the atomic excitation forms a symmetric interference pattern and decays as time evolves. The interference can be interpreted as multiple reflections and transmissions of light in the spin-exchange processes. It is symmetric owing to the reciprocal couplings in the atom-waveguide interface. As the disorder increases, this interference disappears, and a significant amount of the initial excitation sustains when we further raise the disorder strength.

\begin{figure}[t]
\centering
\includegraphics[width=12.0cm,height=6cm]{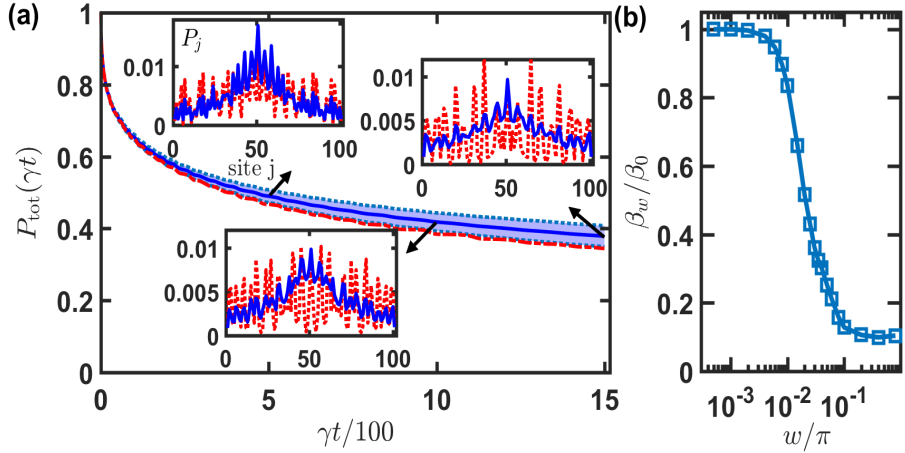}
\caption{Time-evolved localization lengths and power-law scalings of the entropy of entanglement. (a) The total excitation population $P_{\rm tot}(t)$ $=$ $\langle\sum_{j=1}^N P_j(t)\rangle$ (blue-solid line) decays as time evolves for the disorder strength $w/\pi$ $=$ $0.01$. The case without disorder (red-dashed line) is plotted as a comparison, which is close to the lower bound of the shaded area formed by $1\sigma$ deviation (dotted line) of the disordered case. Three insets are population distributions $P_j(t)$ at $\gamma t/100$ $=$ $5$, $10$, $15$, respectively. (b) The fitted power-law scaling of a function $t^{-\beta_{w}}$ in the von Neumann entropy of entanglement for the time range between $\gamma t/100$ $=$ $5-15$. The quotient $\beta_w$ is normalized by the one without disorder and is plotted in logarithmic scales of $w$. For both plots, we consider $N$ $=$ $101$ and $\xi$ $=$ $\pi/8$ as an example.}\label{fig3}
\end{figure}

Next, we use two commonly-used measures of localization lengths and von Neumann entanglement entropy to study the excitation localization demonstrated in figure \ref{fig3}. In figure \ref{fig3}(a), we take one example at a disorder strength when the atomic excitation starts to localize compared to the case without disorder. We can quantify the localization length by fitting $\langle P_j(t)\rangle$ with $e^{-|j-j_c|/j_L}$ where $j_c$ and $j_L$ respectively denote the central site of the atomic chain and a fitted characteristic length. This leads to the full-width-half-maximum $\zeta_L$ $\equiv$ $2j_L\ln(2)$, which can be used as an estimate for the onset of localization when $\zeta_L$ $\lesssim$ $N/2$. We then obtain $\zeta_L$ $=$ $31.6$, $45.6$, and $53.2$ for the three insets in figure \ref{fig2}(a) as time increases. This example shows that $\zeta_L$ changes over time from a localized side ($\zeta_L$ $<$ $N/2$) to a delocalized one ($\zeta_L$ $>$ $N/2$), which, therefore, is not able to provide a deterministic identification of the localization phase transition. 

For von Neumann entanglement entropy $S$, we choose two partitions $S_L$ and $S_R$ with a cut at the center of the chain, which can be calculated as $\langle S_{L(R)}\rangle$ $=$ $\langle{\rm Tr}[\rho_{L(R)}\ln\rho_{L(R)}]\rangle$, where $\rho_{L(R)}$ $\equiv$ ${\rm Tr}_{R(L)}[\rho]$ and $L(R)$ denotes the left (right) of the atomic chain. Under the single excitation space, we obtain, for the example of the left partition, 
\begin{equation}
\rho_L(t) = 
\begin{pmatrix}
|a_1(t)|^2     & a_1(t)a_2^*(t) & \cdots 				 & a_1(t)a_{j_c}^*(t) & a_1(t)a_0^*(t) \\
a_2(t)a_1^*(t) & |a_2(t)|^2     & a_2(t)a_3^*(t) &\cdots  						& a_2(t)a_0^*(t) \\
\vdots 				 & \vdots         & \ddots        &\cdots           		& a_3(t)a_0^*(t) \\
a_{j_c}(t)a_1^*(t) & a_{j_c}(t)a_2^*(t) & \cdots &|a_{j_c}(t)|^2      & a_{j_c}(t)a_0^*(t) \\
a_{0}(t)a_1^*(t) & a_{0}(t)a_2^*(t) & \cdots  &\cdots              & 1-\sum_{l=1}^{j_c}|a_l(t)|^2
\end{pmatrix},
\end{equation}
where $a_0(t)$ $=$ $\sqrt{1-\sum_{j=1}^N |a_j(t)|^2}$ represents the probability amplitude for $\Pi_{j=1}^N |g\rangle_j$. $\rho_L$ is obtained under the reduced subspace of $\sigma_j^\dag|g\rangle^{\otimes j_c}$ with $j$ $=$ $[1,j_c]$ and $\Pi_{j=1}^{j_c} |g\rangle_j$. We note that the trace of $\rho_L$ is unity, which conserves the total probability in this subspace, whereas $\rho_L$ is in general mixed (purity less than one), except at $t$ $=$ $0$, which indicates an entangled composite system once the initial atomic excitation spreads through the whole atomic chain.    

In figure \ref{fig3}(b), we use a power-law fitting $t^{-\beta_w}$ to characterize the trend of the decaying entanglement entropy. The ratio $\beta_w/\beta_0$ reaches one under weak disorders represents the delocalized atomic excitation as the thermalized or the ergodic regime. It drops to a relatively small value and saturates when the disorder strengths are made larger, which corresponds to the localized phase. Similar to the measure of localization lengths, the power-law fitting of entanglement entropy changes depending on the fitting ranges we choose, even though the one we apply in figure \ref{fig3}(b) has covered large enough period. It is this long-time behavior in either $P_{\rm tot}(t)$ or $S_{L(R)}$ that makes the fitted scalings again undeterministic. However, it still provides a qualitative identification on the approximate range of disorder strength that leads to the localized atomic excitation in the atom-waveguide interface. 

\section{Level statistics}

To determine the delocalized to localized atomic excitation, we introduce the gap ratios and their sample variations from the level statistics which is widely used in studying the thermodynamics of quantum chaotic systems \cite{Abanin2019}. The gap ratios $r_j$ can be calculated from the ascendant eigen-energies $E_j$ and the adjacent gaps $\delta_j$ $\equiv$ $E_{j+1}$ $-$ $E_j$, where a dimensionless measure $r_j$ $\equiv$ ${\rm min}\{\delta_j, \delta_{j-1}\}/{\rm max}\{\delta_j, \delta_{j-1}\}$ \cite{Oganesyan2007} lies within $[0,1]$. For each disorder sample, we calculate $r_a$ $\equiv$ $\sum_{j=2}^{N-1}r_j/(N-2)$ and obtain the mean gap ratio $\bar r$ $=$ $\langle r_a\rangle$. Owing to the unique eigenspectrum of the atom-waveguide interface for various dipole-dipole interaction strengths $\xi$ and number of atoms $N$, the level statistics obtained in this particular interface can provide distinctive characteristics of the localization phase in thermodynamics limit.  

From the distribution of spacings between eigen-energies $E_n$, the mean gap ratio $\bar r$ in a tight-binding model of interacting fermions \cite{Oganesyan2007, Sierant2019} has shown the Gaussian orthogonal ensemble (GOE) or Poisson statistics in the nondisordered or disordered conditions, respectively. The GOE presents a level repulsion, $\bar r_{\rm GOE}$ $\approx$ $0.53$, which can be distinguished from the case for Poisson statistics (PS) $\bar r_{\rm PS}$ $\approx$ $0.39$ under a strong disorder. We can also calculate the intrasample variance $\langle v_I\rangle$ $\equiv$ $\langle \sum_{j=2}^{N-2}(r_j^2-r_a^2)\rangle$, which indicates the fluctuations of level repulsions \cite{Sierant2019}. A large level repulsion indicates that the energy level spacings are correlated, which shows the essential feature of the nondisordered system, whereas the disordered one presents a large intrasample variance along with a smaller gap ratio.
  
Since the atom-waveguide system is governed by a non-Hermitian evolution, its eigenspectrum $\bar E_j$ is complex, which involves the eigen-energies Re$[\bar E_j]$ and decay rates Im$[\bar E_j]$. Therefore, the eigen-energies might go with strong decay rates, which deviate from the conventional `eigen-energies' without relaxations. Finite decay rates also correspond to finite lifetimes or spectral broadening, which would effectively smooth out the energy spacings and degrade the definitions of eigen-energies. To remove this issue, we select the valid sectors of energy levels, where their spacings are larger than the average resonance linewidth. We choose (Im$[\bar E_j$ $+$ $\bar E_{j+1}]/2$) $/$ (Re$[\bar E_{j+1}$ $-$ $\bar E_{j}]$) $<$ $0.5$, which is equivalent to specify the patches of eigenspectrum which mostly involves the subradiant sector of decay rates \cite{Jen2020_disorder}.  

\begin{figure}[t]
\centering
\includegraphics[width=12.0cm,height=6cm]{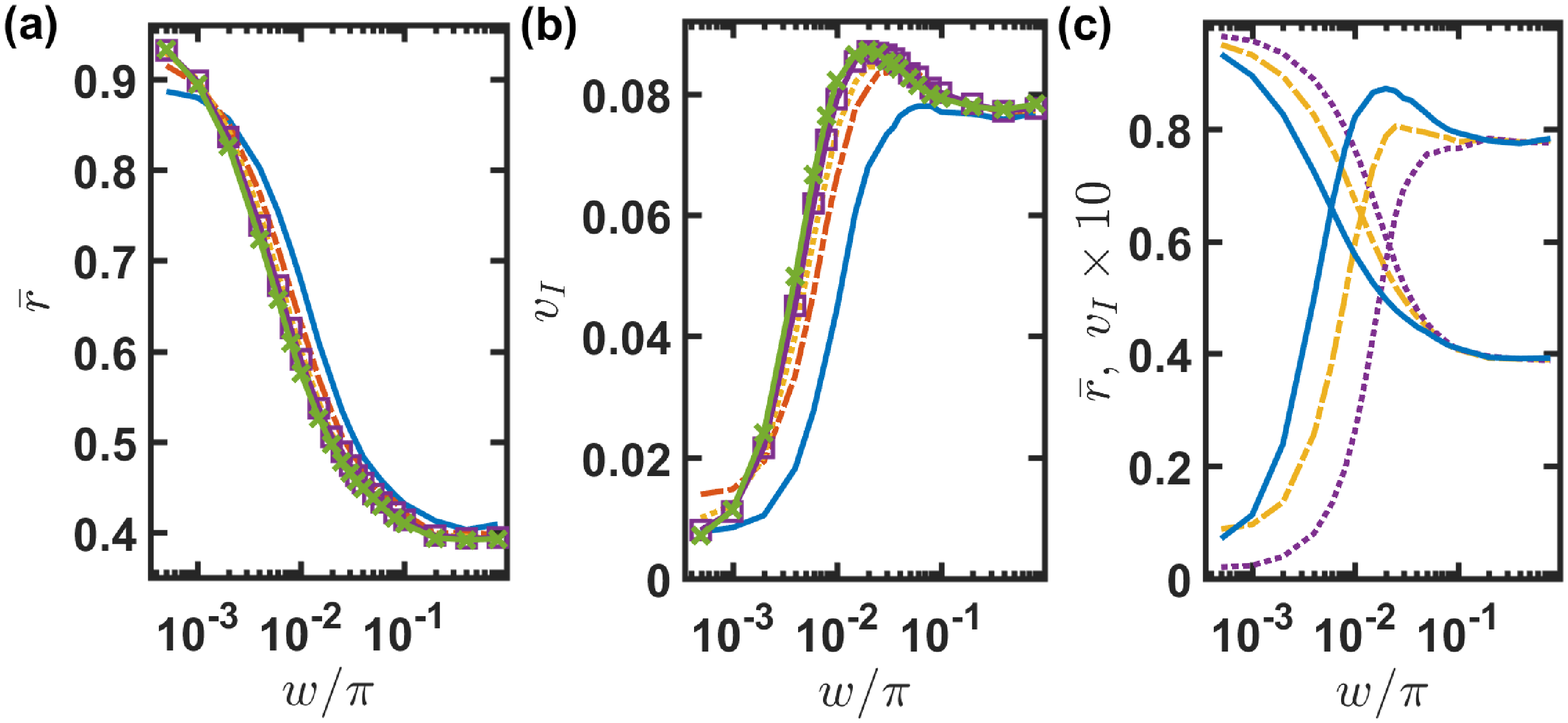}
\caption{The mean and intrasample variance of gap ratios for various disorder strengths. The mean gap ratio (a) decreases while the intrasample variance (b) increases as $w$ increases for $N$ $=$ $101$ (solid line), $201$ (dash-dotted line), $301$ (dotted line), $401$ ($\square$), and $501$ ($\times$) with $\xi/\pi$ $=$ $1/8$. (c) The mean and rescaled intrasample variance of gap ratios for $N$ $=$ $501$ with $\xi/\pi$ $=$ $1/8$ (solid line), $1/4$ (dash-dotted line), and $1/2$ (dotted line) are calculated, which indicates an interaction-driven excitation delocalization at low $w/\pi$ $\lesssim$ $0.02$.}\label{fig4}
\end{figure}

In figures \ref{fig4}(a) and \ref{fig4}(b), we calculate the mean gap ratios and intrasample variance as we increase the number of atoms. An asymptotic collapse of these measures can be seen up to $N$ $=$ $501$, which reaches the thermodynamics limit. As the disorder increases, the mean gap ratio drops from close to one to around $0.4$, which indicates a crossover from the delocalized to the localized phase of an atomic excitation. The mean gap ratio under strong disorder, although its value is close to $\bar r_{\rm PS}$, presents a narrow distribution in huge contrast to Poisson statistics. For a weak disorder, similar narrow distribution appears around the value close to one, resembling a harmonic oscillator spectrum, and this has been found in a Anderson model with a finite size when its localization length is approaching the system size \cite{Torres2019}. In contrast to the mean gap ratios, the intrasample variance increases as the disorder increases owing to the energy level fluctuations in the disordered system. We note that the finite size effect in the trend of the collapse of the curves in figures \ref{fig4}(a) and \ref{fig4}(b) indicates a multiatom enhancement of excitation localization, where at the same disorder strength, the system enters the localization phase much easier for a larger system size.  
 
Finally, in figure \ref{fig4}(c) we plot these two measures of the mean gap ratios and the intrasample variance for different dipole-dipole interaction strengths characterized by $\xi$. The atom-waveguide system with a larger $\xi$ (larger interparticle spacings) has a tendency to favor the delocalization. This is similar to the many-body dipolar systems with long-range interactions \cite{Burin2006, Yao2014, Kucsko2018}, which competes with disorder strengths and thus leads to a facilitation of the delocalized phase. The dominance of disorders over a smaller $\xi$ in the excitation localization can be understood as that the correlated phases among the atoms decohere more easily. By contrast, a larger $\xi$ maintains the long-range correlations which allow the transfer of light excitations at longer distances. We note that at $\xi$ $=$ $0$ or $\pi$, the decoherence-free eigenmodes are sustained in the system. Therefore, the atomic excitation decays more rapidly under disorders than without, which is opposite to the localization phenomenon under disorders and dissipations investigated here.  

\section{Discussion and conclusion}

The atom-waveguide interface presents one of the strong coupling platforms that can store and relay quantum states of light. One of the advantages in this interface is the scalability and controllability that are essential in quantum network \cite{Kimble2008}. In addition to its potential application in quantum storage and quantum communication, the localized atomic excitation corresponds to the retained memory of the initialized states, which offers another opportunity to preserve quantum information. Many different platforms can realize this quantum interface, including quantum dots, superconducting qubits, neutral atoms, or Rydberg atoms, while a strong coupling regime in respective light-matter interacting systems is required for efficient operations to remove the effect of nonradiative or external losses. A strong coupling regime can be approached using quantum dots on the nanofiber \cite{Yala2014} or assisted with an external cavity, for example. 

In the atom-waveguide interface with long-range dipole-dipole interactions under dissipations, we have studied the characteristic localization lengths and entanglement entropy, which can qualitatively indicate the onset of excitation localization. We further use the level statistics, extracted from the eigenspectrum of the underlying non-Hermitian Hamiltonian, to clearly identify the crossover from the delocalization to localization phases. From the level statistics, the localized (delocalized) phase shows small (close to one) and significant (insignificant) values of the mean gap ratio and intrasample variance, respectively, which is a distinctive crossover behavior in the atom-waveguide interface. Our results can offer insights to the nonequilibrium dynamics with long-range couplings and pave the way toward understanding many-body localized systems under dissipations. One potential extension to our study can be investigating long-time spin dynamics after multiple spin flips, where long-range spin-spin correlations would emerge and rich nonequilibrium dynamics could arise in the many-body interacting platforms.


\ack
We thank C.-Y. Lin and M.-S. Chang for insightful discussions on the dipolar interacting system. This work is supported by the Ministry of Science and Technology (MOST), Taiwan, under the Grant No. MOST-109-2112-M-001-035-MY3.

\end{document}